\documentstyle[12pt]{article}
\begin{document}
\baselineskip=25pt
\begin{flushright}
Preprint Dipartimento di Fisica dell'Universit\`a di Catania,\\
rif. 95/03
\end{flushright}

\begin{center}
\bf Building scars for integrable systems
\end{center}

\begin{center}
\normalsize\em M. Baldo$^{1}$, F. Raciti$^{2}$\\
$^1$INFN, Sezione di Catania, 57 Corso Italia , I-95129 Catania,
Italy\\
$^2$ Dipartimento di Fisica and INFN, 57 Corso Italia,
 I-95129 Catania, Italy
\end{center}

\vskip 3 true cm

\begin{abstract}
It is shown, by means of a simple specific example,
that for integrable systems it is possible to build up approximate
eigenfunctions, called {\it asymptotic eigenfunctions}, which are
concentrated as much as one wants to a classical trajectory and
have a lifetime as long as one wants. These states are directly related
to the presence of shell structures in the quantal spectrum of the system.
It is argued that the result can be extended to classically chaotic system,
at least in the asymptotic regime.
\end{abstract}

\newpage

{\bf 1. Introduction}
\vskip .2 true cm
\par
The problem of quantising classically chaotic systems has attracted much
attention in the last few years, especially in view of the subtle
connection between classical and quantum mechanics in the semi-classical
limit, generally referred as the $\hbar \rightarrow 0$ limit. Besides the
well studied spectral properties of quantal
systems, whose classical counterpart is chaotic, one of the most interesting
and intriguing phenomenon is the appearance of the so called ``scars",
i.e. eigenfunctions which display a strong concentration of probability
along a classical closed trajectory. This phenomenon is not yet well
understood, and it has been observed in billiards as well as in smooth
hamiltonian systems. It has been argued that the scars phenomenon can be
directly related to the presence of  a peak structure in the
spectral density $^{[1],[2]}$, which seems to favour a particular stability
of a wave-packet moving along a classically unstable trajectory. This problem
has been considered in ref.$[3]$ where it has been suggested that
scars structures could be revealed through an energy averaging, in the
framework of the Gutzwiller$^{[4]}$ formulation of the semi-classical
limit.
In order to shade some light on this problem, we consider the
problem of establishing to what extent, for an {\it integrable} system,
it is possible to construct an (approximate) eigenfuction which is
close as much as possible to a classical trajectory. In the short
wavelength limit, Erhenferst theorem assures that the motion of a wave
packet can be as close as one wants to a classical trajectory, but,
of course, an eigenfunction has to be time independent. Therefore,
the motion of the wave packet in the semi-classical limit can only
eventually suggest the particular superposition of wave packets which
is (almost) time independent and still concentrated around the trajectory.
However, to estimate the rate of
spreading of a wave packet in the presence of a potential or a barrier
is not an easy task, and can be performed only numerically, even for
an integrable system. In the case of an integrable system,
one can try to build up directly a superposition of exact eigenstates
in order to construct
a wave packet which is concentrated along some classical trajectory and
at the same time survives as long as one can. The essential point of
the problem is to recognize that the
lifetime of such a wave packet must be much longer than the
characteristic time of the classical motion, namely the period of the
trajectory, otherwise the wave packet cannot be considered an
approximate eigenstate to any respect. This condition is {\em not}
fulfilled in the treatment of ref.$[3]$  for chaotic systems.
We will show that in the case of integrable systems this condition can be
fulfilled to any degree of precision, by building up a suitable
superposition of eigenstates which are approximately degenerate,
namely they belong to the same ``shell" of the spectrum. It is hoped
that this result could be of some help of solving the scars problem
also in the case of chaotic systems, where the appearance of a peak
in the density of state can be viewed as a sort of an  ``accidental"
(approximate) shell. Actually, the wave functions belonging
to this class, being
strongly localized, are sensitive only to the local shape of the
billiard. Therefore, their connection with classical
orbits is not affected if the billiard shape is distorted
along the contour where the wave function is essentially zero.
This can include the chaotic
cases, at least for asymptotically large quantum numbers.

\

\leftline{\bf 2. Method}
\vskip .2 true cm
\par
Let us consider a very simple integrable system with two degrees of
freedom, the circular billiard. The system is trivially solvable,
the eigenfunctions are cylindrical Bessel functions
$ J_l(\rho_{nl}r/R)\exp (i l\phi) $,
and the $\rho_{nl}$ 's are the zeroes of the Bessel fuctions (BF).
Here $R$ is the billiard radius, r the radial coordinate and $\phi$
the angular coordinate. The quantum
numbers $n, l$ correspond to the quantisation of the radial and angular
motion of the particle respectively. This can be seen in the semi-classical
limit, namely for large values of the quantum numbers, for which the
asymptotic form of the BF${[5]}$ gives the semi-classical
(Bhor-Sommerfeld) quantisation condition
$$
 \sqrt{k_{nl}^2 R^{2} - l^2} - l\beta_0 \, =\, ( 2n + 1 )\pi/2 + \pi/4
\eqno(1)
$$
\noindent
and it is readly verified that right hand side is just the action
integral along the classical radial motion.
For future considerations,
it is convinient to keep in mind that the (constant) angular
distance $2\beta_0$ between two successive hits of the particle at
the billiard wall is given by $\cos (\beta_0) = l/(k_{nl}R)$.
Shell structures are
associated with closed orbits $^{6}$. In fact, the condition of
$k_{nl}$ being stationary for variations $\Delta n$ and
$\Delta l$ of the quantum numbers, from eq. (1), reads
$$
{\beta_0 \over \pi} = l_0/(\rho_{n_0l_0}) \, =\,
{{\Delta n}\over {\Delta l}} = {p \over q}
\eqno(2)
$$
\noindent
being $p$ and $q > p$ two integer numbers with no common prime factor.
This implies that the
corresponding classical trajectory closes after $m = pq$ hits at the wall,
and the quantum numbers $n$ and $l$ are linearly related. Eq. (2) imposes
also a condition for the quantum numbers $l_0$ and $n_0$, which
asymptotically can be satisfied with arbitrary precision.
The states which have quantum numbers $l = l_0 + \Delta l $
and $ n = n_0 + \Delta n$ satisfying
the linear condition of eq. (2) are approximately degenerate, and therefore
form an energy shell, which, in turn, implies the appearance of a sharp
peak in the density of states. We want to show now, by elementary
considerations, that for large enough quantum numbers $n_0$ and $l_0$
one can construct linear superpositions of the eigenstates belonging
to the same shell, which satisfy {\em at the same time} the
two conditions, a) they are concentrated with arbitrary precision
along a classical (closed) trajectory b) they have a lifetime
arbitrarly longer than the classical characteristic time
$T = M/\rho_{nl}{\hbar}$, being $M$ the mass of the particle. Because of
the simple symmetry of the system, each eigenfunction
has a rotationally invariant probability density. A wave packet
with a narrow angular spread $\Delta \phi = 1/\Delta_l$ can be written
$$
 \Psi_{l_0}(r,\phi) \, =\, \sum_{nl} \exp\left( {{(l - l_0)^2} \over
  {2\Delta_l^2}}\right)  \exp(il\phi) J_l (k_{nl}r)
\eqno(3)
$$
\noindent
In order to minimize the energy spread, we restrict the summation
to the eigenstates belonging to the same shell. Asymptotically,
for large quantum numbers, the summation can be performed by expanding
the phase of the BF around the chosen value $l_0$ of the angular
momentum and taking into account the condition of stationary value
of $k_{nl}$ and the corresponding linear relation between the
quantum numbers $n$ and $l$. The result reads
$$
 \Psi_{l_0}(r,\phi) \,\, \sim \,\,  \exp\left(
  {{(\phi - \beta(r))^2}\over {2\Delta \phi^2}} \right)
\eqno(4)
$$
\noindent
where $\cos (\beta(r)) = l_0/(k_{n_0l_0}r) \,$ , and $\, k_{n_0l_0} =
\rho_{n_0l_0}/R$ is the eigenmomentum. The wave packet of eq. (3)
is clearly concentrated around the classical trajectory,
which is a polygon or a star with $m$ sites, with a spatial spread
$\Delta s \sim R \Delta \phi \sim R/\Delta_l$.
The energy spread $\Delta E$ can be estimated in terms of the
second derivative $k^{\prime\prime}$ of $k_{nl}$ at $n_0 l_0$ along
the direction defined by eq. (2), $\Delta E/ E \, =\,
(k^{\prime\prime} / k_{nl}) \Delta_l^2 $. After some algebra, one gets
$\Delta E/E \, =\, g(\Delta_l/\rho_{n_0l_0})^2$, where $g$ is a
costant factor of order unity, and it can be checked that the
higher order terms are vanishing small for asymptotic quantum numbers.
The ratio between the quantal lifetime
$\tau_q =\hbar /\Delta E$ and the classical characteristic time $T$ turns out
$$
 {\tau_q \over T} \, \sim \, \left({{\Delta s} \over R}\right)^2 l_0
\eqno(5)
$$
\noindent
This ratio, for a fixed value of the localization $\Delta s$, can be
made arbitrarly large by increasing the values of the quantum numbers.
Had we chosen a different linear combination, on the contrary this ratio
would have been of order unity. In other words, the uncertainity $\Delta E$,
with the contraint of eq. (2), is asymptotically of the same order of
the mean level spacing $n(E) \, =\, 2\pi MR^2/\hbar^2$. An example,
corresponding to $ p = 1$ and $ q = 3$ is depicted

\

\centerline{\large \bf Table 1}

\

\centerline{\begin{tabular}{ccc}\hline\hline
\multicolumn{1}{c}{$\quad\qquad\qquad$}&\multicolumn{1}{c}{$\quad\qquad\qquad$}&
\multicolumn{1}{c}{$\quad\qquad\qquad$}\\
$l_0$                                &   $n_0$
                           & $\rho$\\
                                     &
                           &  \\ \hline
                                     &                                     &
\\
111                                  &  30
                           & 241.87\\
114                                  &  29
                           & 242.00\\
117                                  &  28
                           & 242.09 \\
120                                  &  27
                           & 242.14 \\
123                                  &  26
                           &242.13 \\
126                                  &  25
                           &242.07\\
129                                  &  24
                           & 241.96\\
                                  &                                     & \\
\hline\hline
\end{tabular}}

\

\

in Fig. 1. The wave
function is calculated with the expression of eq. (4) and with the full
expansion of eq. (3) in part $a$ and $b$ respectively. Table 1 reports
the quantum numbers and energies used in the calculation. One can notice
that a high degree of localization can be obtained with only few terms
and not too large quantum numbers. In the example $\tau_q / T \, \sim
\, 15$. Higher degree of localization and longer lifetime can be obtained
following the above described procedure.

\newpage

\leftline{\bf 3. Discussion and conclusions}
\vskip .2 true cm
\par
The wave function depicted in Fig. 1 has a striking similarity with some
scars reported in ref. $[7]$ for a chaotic billiard. The authors
show that this type of scars ``live" in thin invariant tori embedded
in a chaotic region. In the procedure of the present paper the integrable
billiard can be deformed in regions of the contour where the wave function
is vanishing small, and the billiard could become of the
chaotic type leaving the scars essentially untouched. This could be a
mechanism of generating scars in a chaotic system. Of course, the
procedure does not exhaust all the possibilities, and indeed in ref.
$[7]$ many examples are shown where the scars ``live" in the
classically chaotic region (in the Wigner tranform sense).
 Another possibility of generalizing the procedure to chaotic billiard
is the case of ``local integrability", described in details in ref.
$[8]$. In this case the presence of an adiabatic barrier allows
to expand the hamiltonian, around a closed trajectory, in the longitudinal
and transverse actions, which are approximate constants of the motion in the
vicinity of the trajectory. In this case the present procedure can be
repeated step by step, substituting the angular momentum with the transverse
action and the angular coordinate with the transverse coordinate. \par

In conclusion we have presented a procedure to build up scars for integrable
systems. Generalizing the method to chaotic systems appears possible,
at least when thin invariant tori exist or local integrability
is present. The extension of the method to more generic cases is
under study and the results will be reported elsewhere.

\vfill\eject

\centerline{\bf  References }
\begin{itemize}
\item{[1]}  Eric.J. Heller,{\em Phys.Rev.Lett.} {\bf 53} 1515 (1994)
\item{[2]} G.G. de Polavieja, F. Borondo, R.M.Benito,
{\em Phys.Rev.Lett.} {\bf 73} 1613  (1994)
\item{[3]} E.B.Bogomolny, {\em Physica} {\bf D 31} 169 (1988)
\item{[4]} M.C.Gutzwiller, {\em Chaos in classical and quantum mechanics},
Springer-Verlag, 1991
\item{[5]} I.S.Gradshteyn, I.M.Ryzhik, {\em Table of Integrals, series and
products}, Academic Press,INC, 1991, formula 8.453
\item{[6]} R.Balian and R.Bloch,{\em Ann.Phys} {\bf 69} 76 1971;

A.Bohr, B.R.Mottelson,{\em Nuclear structure}, vol.{\bf II},
W.A.Benjamin, Inc. (1975), p. 579 and ff.
\item{[7]} B.Li and M.Robnik, {\em J.Phys} {\bf A 28} 2779 (1995)
\item{[8]} B. Eckhardt et al., {\em Phys.Rev} {\bf A 39} 3776 (1989)
\end{itemize}
\bigskip
\vfill\eject
\centerline{Figure captions}
\begin{itemize}
\item{1.a} Contour plot of the probability density associated to the
wave function (3). Fixing an angular spread  $\Delta \varphi=0.25$,
and considering the contribution of only 6 values of $l$ centered
around $l_0=120$ (cfr. table 1), we have obtained a ratio
$\frac {\tau_q}{T}=15$

\

\item {1.b}  Contour plot of the probability density  associated
to the wave function (4)
 which is obtained from (3)  expanding the phase
of the BF  around $l_0$  taking into account the condition
of stationary value of $k_nl$ and replacing
the summation in (3)
by an integral over $l$.

\end{itemize}
\end{document}